\RequirePackage{fix-cm}
\documentclass[twocolumn,epjc3]{svjour3}  
\smartqed  
\RequirePackage{graphicx}
\usepackage{graphics,epsfig}
\usepackage{subfigure}
\usepackage{float}
\usepackage{epstopdf}
\usepackage{graphicx}
\usepackage{dcolumn}
\usepackage{amsmath}
\usepackage{epstopdf}
\usepackage{enumitem}
\usepackage{lmodern}
\usepackage{fix-cm}

\journalname{Noname}
\begin{document}

\title{Geodesics and Light Deflection in Schwarzschild-like Spacetime from Cosmology-Inspired Modified Gravity
}

\author{
Ritesh Pandey\thanksref{e1,addr1}
\and
Shubham Kala\thanksref{e2,addr2}
\and
Amare Abebe\thanksref{e3,addr1,addr3}
\and
Hemwati Nandan\thanksref{e4,addr4,addr1}
\and
G.G.L. Nashed\thanksref{e5,addr5,addr1}
}

\thankstext{e1}{e-mail: riteshphy0@gmail.com}
\thankstext{e2}{e-mail: shubhamkala871@gmail.com}
\thankstext{e3}{e-mail: amare.abebe@nithecs.ac.za}
\thankstext{e4}{e-mail: hnandan@associates.iucaa.in}
\thankstext{e5}{e-mail: nashed@bue.edu.eg}

\institute{Centre for Space Research, North-West University, Potchefstroom 2520, South Africa \label{addr1} \and
The Institute of Mathematical Sciences, C.I.T Campus, Taramani-600113, Chennai, India \label{addr2} \and
National Institute for Theoretical and Computational Sciences (NITheCS), South Africa \label{addr3} \and
Department of Physics, Hemvati Nandan Bahuguna Garhwal Central University, Srinagar-246174, Uttarakhand, India \label{addr4} \and
Centre for Theoretical Physics, The British University in Egypt, Cairo 11837, Egypt \label{addr5} }

\date{Received: date / Accepted: date}

\maketitle

\begin{abstract}
We investigate cosmology-driven modifications to Schwarzschild-like black hole spacetimes and analyze their impact on photon propagation, gravitational lensing, and shadow observation. The gravitational deflection angle is computed using the Rindler–Is-hak method, which incorporates finite-distance corrections and provides a consistent framework for non--asym-ptotically flat spacetimes. The effective potential for null geodesics exhibits a single unstable maximum corresponding to the photon sphere, and we study photon orbits classified according to the critical impact parameter into capture, escape, and unstable circular trajectories. Our analysis shows that the deflection angle decreases with increasing model parameter $(\alpha)$, resulting in weaker light bending compared to the Schwarzschild case. In addition, we examine the angular diameter of the black hole shadow as measured by a static observer, highlighting its dependence on the cosmological modification parameters. These results suggest that high-precision astrometric and lensing observations can place meaningful constraints on cosmology-inspired modifications to gravity, thereby linking astrophysical black holes with cosmic expansion and offering a novel probe of gravitational physics in strong-field regimes.
\end{abstract}

\section{Introduction}
\noindent The latest data from Type Ia supernovae (SNe Ia), the cosmic microwave background,
baryon acoustic oscillations (BAO), and direct measurements of the Hubble parameter indicate that the universe is currently undergoing an accelerated phase of expansion~\cite{SupernovaSearchTeam:1998fmf,Chevallier:2000qy}. For instance, model-independent BAO reconstructions using Dark Energy Spectroscopic Instrument (DESI) data confirm continued acceleration up to  $z \approx 0.7$ at nearly 99.7\% confidence~\cite{DESI:2025wyn}, while the new Union3 supernova sample hints at possible dynamical behavior of dark energy beyond the standard $\Lambda$CDM picture~\cite{Chudaykin:2024gol}.
 This points to the reality of an enigmatic form of energy that is possessed with a very negative pressure, widely known as dark energy~\cite{Padmanabhan:2004av,Straumann:2006tv}. Finding the true nature of this mysterious component and understanding its implications for cosmology has emerged as one of the most challenging problems of modern cosmology and astrophysics. An alternative perspective is that the current acceleration of the universe might result not from some exotic dark energy supplement but rather from some alteration of gravity at sub-horizon scales. Dvali-Gabadadze-Porrati (DGP) brane-induced gravity~\cite{dvali20004d}, the Cardassian model~\cite{freese2002cardassian}, and the Dvali-Turner model~\cite{dvali:2003rk} are some which produce late-time acceleration by modifying the Friedmann equation without outright dark energy. Within this perspective, the observed expansion is, to a certain degree, self-consistent and can be derived from the matter content. The geodesic structure of spherically symmetric spacetime governs the evolution. It is possible to determine the modified gravity theory by setting the scale factor to a dust-dominated universe and applying a generalized Birkhoff’s law~\cite{lue2004differentiating,barreiro2004generalized}. The approach of deriving the Schwarzschild-like solution of modified gravity \cite{lue2004differentiating} that produces the desired geometric expansion of cosmology, assuming a universe filled with dust, has been discussed in detail by Lue et al~\cite{lue2004differentiating}.

Matter and energy modify the form and curvature of spacetime in a fundamental way in General Relativity (GR)~\cite{hartle2021gravity,poisson2004relativist,Wald:1984rg}. Astrophysics has long relied on geodesic motion as a powerful tool to probe different gravitational backgrounds~\cite{chandrasekhar1998mathematical}, since geodesics describe the trajectories of free-falling particles and light rays \cite{Breton:2002td,hackmann2008geodesic,Grunau:2010gd,hackmann2010geodesic,Abbas:2014oua,uniyal2015geodesic,Soroushfar:2016yea,Kala:2021ppi,Battista:2022krl,Capozziello:2025wwl}. In this article, we explore the generalized Chaplygin gas (GCG) model within the framework of a modified gravity approach, following the ideas of Lue et al ~\cite{lue2004differentiating}. Rather than introducing an exotic energy density that obeys the GCG equation of state to explain cosmic acceleration, we assume that gravity itself is modified to produce the observed expansion~\cite{lue2004differentiating,barreiro2004generalized}. We require that the cosmological background evolution remains compatible with the GCG model, with the caveat that the energy density consists only of matter while neglecting the negligible contribution of present-day radiation. Under these assumptions, we derive a Schwarzschild-like metric consistent with the underlying modified gravity theory responsible for the assumed expansion history. We then analyze the geodesic motion in this spacetime through the effective potential and orbital structure to gain insight into the effects of the modified geometry.

Gravitational lensing, the bending of light rays in curved spacetime, serves as a powerful probe of gravity in both weak and strong-field regimes \cite{Virbhadra:1999nm,Bozza:2001xd,Iyer:2006cn,Cunha:2018acu,Kala:2020prt,Kala:2020viz,Kala:2022uog,Kala:2024fvg,Kala:2025fld,Roy:2025hdw,Roy:2025qmx,Kala:2025iri,Kukreti:2025rzn,Feleppa:2024kio,Feleppa:2024vdk,Ahmed:2023dvc,Ahmed:2024fye,Pantig:2024lpg,Vishvakarma:2024icz,ahmed2025gravitational,Waseem:2025yib,Lambiase:2024uzy,Turakhonov:2025ojy}. In scenarios motivated by Cosmology-Inspired Modified Gravity, the deflection angle is influenced not only by the local gravitational potential of the lensing object but also by large-scale cosmological corrections that modify the underlying metric \cite{Blandford:1991xc,Ratra:1992ut}. Such deviations can imprint observable signatures on lensing phenomena, including shifts in image positions, changes in magnification patterns, and modifications to the structure of relativistic images near the photon sphere \cite{Bacon:2000sy,Hoekstra:2008db}. These effects provide a unique opportunity to test extensions of General Relativity using high-precision astrophysical observations, ranging from the Event Horizon Telescope’s imaging of supermassive black holes to forthcoming space-based gravitational lensing surveys. Gravitational lensing has been extensively studied as a tool to probe both the strong and weak-field predictions of gravity, providing constraints on modified gravity models through precise measurements of light deflection \cite{Aghili:2014aga,Islam:2021ful,Kumar:2021cyl,Wang:2024iwt}.

Supermassive black holes residing at galactic centers, such as Sagittarius A* in the Milky Way, are predicted to produce a shadow corresponding to the locus of photon trajectories asymptotically approaching unstable circular orbits at the photon sphere. The angular size of this shadow, measurable by a static observer, is determined by the critical impact parameter and hence encodes the structure of the underlying spacetime geometry. In the classical Schwarzschild case, the shadow size is uniquely fixed by the horizon and photon sphere radii~\cite{Synge:1966okc}, whereas in Schwarzschild-like modifications that include a Hubble parameter $(H_{0})$ or cosmological constant $(\Lambda)$, the lapse function acquires cosmological-scale corrections that shift the photon sphere location and consequently alter the shadow radius~\cite{Perlick:2018iye,Bisnovatyi-Kogan:2018vxl,Bisnovatyi-Kogan:2019wdd}. Such deviations are of particular relevance in the context of recent observational constraints from the Event Horizon Telescope (EHT)~\cite{EventHorizonTelescope:2019dse}, as well as extensive theoretical investigations on modified gravity and black hole shadow properties~\cite{Li:2020drn,Cotaescu:2020kcr,Cuzinatto:2023kbo}.

The organization of the paper is as follows: In Section~\ref{Sec2}, we introduce the Schwarzschild-like spacetime in Cosmology-Inspired Modified Gravity, along with a brief discussion of the horizon structure. In Section~\ref{Sec3}, we derive the equations of motion around this black hole spacetime. Section \ref{Sec4} is devoted to the study of the effective potential and possible types of orbit using null geodesics. In Section ~\ref{Sec5}, we obtain an approximate expression in the weak-field limit and compare the results with the classical GR prediction. In Section~\ref{Sec6}, we analyze the angular radius of the black hole shadow as measured by a static observer, highlighting how it varies with observer distance and spacetime modifications. Finally, we summarize and conclude our findings in Section~\ref{Sec7}. In this work, we set $G = c = 1$ and measure the lengths in geometric units of the black hole mass ($M = 1$). The physical Hubble constant $H_0^{\rm phys}$ (in s$^{-1}$) is converted to a dimensionless form by $H_0 = H_0^{\rm phys} \, (G M / c^3)$. A brief description of this conversion procedure is provided in Appendix~A. Unless otherwise stated, all dimensionless parameters and quantities are expressed in these geometric units.
\section{Schwarzschild-like Spacetime in Cosmology-Inspired Modified Gravity} \label{Sec2}

Investigating gravitational interactions on cosmos and astrophysical scales gives a chance to test possible modifications to GR. GR is well tested within the solar system and with binary pulsar systems, while the ever-increasing expansion of the Universe has led to the consideration of modification of gravity theories capable of reproducing the observed acceleration of the universe, and without the need to assume dark energy. We investigate a class of spherically symmetric Schwarzschild-like metrics that appear due to modification of the Einstein equations where the equation of state of dust matter is kept intact but the Friedmann equation is modified at later epochs. We study the conditions under which these models converge to GR at small scales and study their astrophysical consequences.
Following the procedure described in \cite{lue2004differentiating}, it is attainable to obtain the Schwarzschild-like metric of this modified gravity around a spherically symmetric matter source. In general the metric is given by,
\begin{equation}
  ds^2 = -f(r) \, c^2 dt^2 + \frac{dr^2}{f(r)} 
+ r^2 \left( d\theta^2 + \sin^2\theta \, d\phi^2 \right),  
\end{equation}
where \(f(r)\) is 
\begin{equation}
f(r) = 1 - r^2 H_0^2 \, g\!\left( \frac{r_c^3}{r^3} \right),    
\end{equation}
with
\begin{equation}
 g(x) = \left[ \left(1 - \Omega_m^{\alpha+1} \right) + x^{\alpha+1} \right]^{\frac{1}{\alpha+1}}.   
\end{equation}
Here, $H_0$ denotes the Hubble constant, $r_c$ is the crossover scale controlling the strength of the modification, $\alpha$ is the model parameter determining the deviation from the Schwarzschild solution, and $\Omega_m$ represents the current matter density parameter. The parameter \(x\) is a dimensionless quantity
$x = \frac{r_c^3}{r^3}.$ with $r_c = \left( \frac{2GM}{H_0^2} \right)^{1/3}$.
 In the early universe, the conventional Einstein equation is recovered for \(r \ll r_c\), we have \(x \gg 1\)  \(g(x) \to x\),  but in the late universe, gravity is altered. It is intriguing to consider the potential consequences of modifying Einstein's gravity at astrophysical scales, given that we assume that this modification is the cause of the scale factor's development.  So for condition \(g(x) \to x\), we can recover f(r) for the Schwarzschild metric.
\[
f(r) \approx 1 - \frac{2GM}{r},
\]
\begin{figure}[htbp]
    \centering
    \includegraphics[width=0.5\textwidth,height=0.35\textheight]{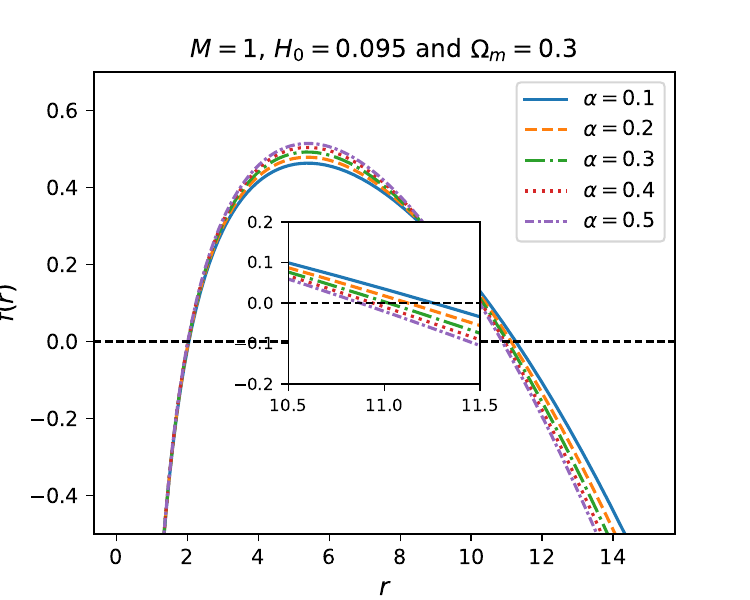}
    \caption{The variation of lapse function $f(r)$ with radial distance ($r$) for different values of $\alpha$.}
    \label{fig:frModSBHalpha}
\end{figure}

In Fig.~\ref{fig:frModSBHalpha}, we plot the lapse function \( f(r) \) for different values of the parameter \(\alpha\) in order to examine the formation of horizons. For fixed \( H_{0} = 0.095 \) and \(\Omega_{m} = 0.3\), the plots reveal that the spacetime possesses two distinct horizons: an inner black hole event horizon and an outer cosmological horizon across the entire considered range of \(\alpha\). This behavior indicates that the parameter \(\alpha\) does not induce a horizon-merging scenario that could lead to an extremal configuration, nor does it eliminate either of the horizons entirely. From a physical standpoint, the persistence of both horizons suggests that the underlying spacetime structure remains that of a non-extremal black hole embedded in a cosmological background for all tested \(\alpha\) values, implying that modifications introduced via \(\alpha\) primarily shift the radial positions of the horizons without altering their qualitative nature.

\section{Equations of Motion in Cosmology-Inspired Modified Gravity} \label{Sec3}
We consider a static, spherically symmetric line element
\begin{equation}
\label{eq:metric}
ds^2 = -f(r)\,dt^2 + \frac{dr^2}{f(r)} + r^2\!\left(d\theta^2 + \sin^2\!\theta\, d\phi^2\right),
\end{equation}
and study test-particle (timelike) and photon (null) motion via the Lagrangian
\begin{equation}
\label{eq:Lagrangian}
\mathcal{L}=\tfrac12\, g_{\mu\nu}\dot{x}^\mu\dot{x}^\nu
=\tfrac12\!\left[-f(r)\dot t^{\,2}+\frac{\dot r^{\,2}}{f(r)}+r^2\dot\theta^{\,2}+r^2\sin^2\!\theta\,\dot\phi^{\,2}\right],
\end{equation}
where a dot denotes differentiation with respect to an affine parameter $\lambda$. The Killing vectors $\partial_t$ and $\partial_\phi$ imply two conserved quantities,
\begin{equation}
\label{eq:EL_defs}
E \equiv -\frac{\partial\mathcal{L}}{\partial \dot t}=f(r)\,\dot t,
\qquad
L \equiv \frac{\partial\mathcal{L}}{\partial \dot \phi}=r^2\sin^2\!\theta\,\dot\phi,
\end{equation}
interpreted as the (specific) energy and angular momentum. By spherical symmetry we set initial conditions in the equatorial plane; then $\theta=\pi/2$ and $\dot\theta=0$ hold along the geodesic, so $L=r^2\dot\phi$.

The normalization of the four-velocity gives the first integral
\begin{equation}
\label{eq:norm}
g_{\mu\nu}\dot x^\mu\dot x^\nu=\epsilon,\qquad
\epsilon=\begin{cases}
1 & \text{timelike geodesics},\\[2pt]
0 & \text{null geodesics},
\end{cases}
\end{equation}
which, upon substituting \eqref{eq:EL_defs}, yields the radial equation
\begin{equation}
\label{eq:radial_first_integral}
\dot r^{\,2}=E^2 - f(r)\!\left(\epsilon+\frac{L^2}{r^2}\right)
\;\equiv\; E^2 - V_{\rm eff}(r;L,\epsilon),
\end{equation}
with the effective potential
\begin{equation}
\label{eq:Veff}
V_{\rm eff}(r;L,\epsilon)=f(r)\!\left(\epsilon+\frac{L^2}{r^2}\right).
\end{equation}
Turning points satisfy $E^2=V_{\rm eff}$; circular orbits additionally require $V_{\rm eff}'(r)=0$, with stability determined by the sign of $V_{\rm eff}''(r)$.

In the following, we restrict our analysis to null geodesics ($\epsilon=0$). Defining the impact parameter $b\equiv L/E$, Eq.~\ref{eq:radial_first_integral} becomes
\begin{equation}
\label{eq:null_radial_b}
\dot r^{\,2}=E^2\!\left[\,1-\frac{f(r)\,b^2}{r^2}\right],
\end{equation}
or, eliminating the affine parameter in favor of $\phi$,
\begin{equation}
\label{eq:drdphi_form}
\left(\frac{dr}{d\phi}\right)^{\!2}
=\frac{r^4}{L^2}\,\dot r^{\,2}
=\frac{r^4}{b^2}-r^2 f(r).
\end{equation}
This is the equation of motion for photons, which depends explicitly on the metric function and the model parameters. By solving this equation, we can determine the possible types of photon orbits around the black hole. In the next section, we investigate these trajectories in detail and analyze how the spacetime geometry influences their properties.

\section{Effective Potential and Possible types of Orbits} \label{Sec4}
\begin{figure}[H]
    \centering
    \includegraphics[width=0.45\textwidth,height=0.28\textheight]{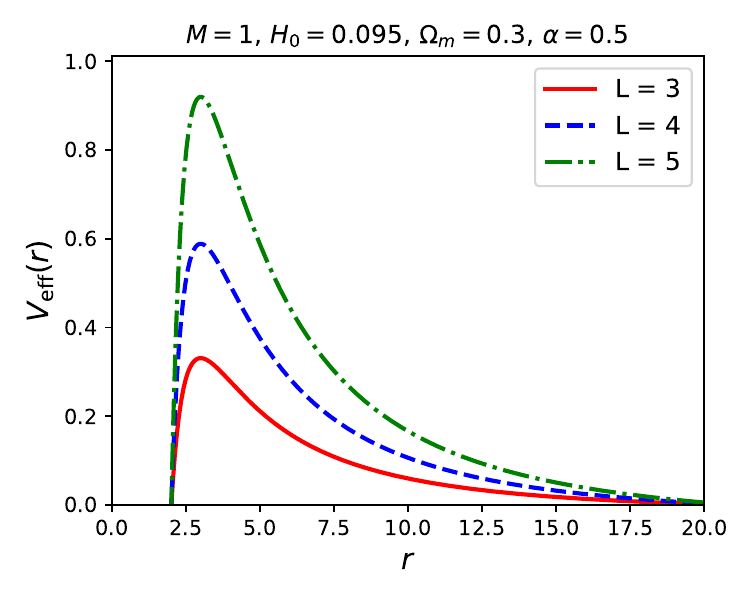}
    \caption{The variation of effective potential with radial distance for different values of angular momentum of massless particles.}
    \label{fig:VeffModSBH1}
\end{figure}

\begin{figure}[H]
    \centering
    \includegraphics[width=0.45\textwidth,height=0.28\textheight]{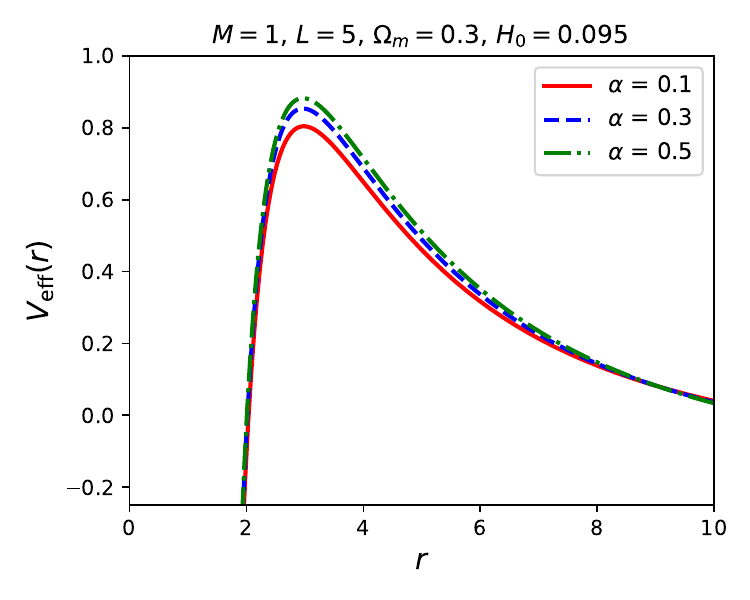}
    \caption{The variation of effective potential with radial distance for different values of Hubble length parameter.}
    \label{fig:VeffFig1}
\end{figure}

The effective potential formalism provides a convenient way to analyze the motion of test particles and photons in a given spacetime geometry. By studying its shape, one can identify possible circular orbits, stability conditions, and the existence of turning points in the geodesic trajectories. By setting $\epsilon=0$ in Eq.~\ref{eq:Veff}, the effective potential for null geodesics can be written as
\begin{equation}
V_{\mathrm{eff}}(r) = f(r) \left( \frac{L^2}{r^2} \right),
\end{equation}
where
\begin{equation}
f(r) = 1 - r^{2} H_{0}^{2} \left[ \left( 1 - \Omega_{m}^{\alpha+1} \right) + \left( \frac{r_{c}^{3}}{r^{3}} \right)^{\alpha+1} \right]^{\frac{1}{\alpha+1}},
\end{equation}
and
\begin{equation}
r_{c} = \left( \frac{2GM}{H_{0}^{2}} \right)^{1/3}.
\end{equation}
The variation of the effective potential with radial distance is shown in Fig.~\ref{fig:VeffModSBH1} for different values of the angular momentum of a massless particle, and in Fig.~\ref{fig:VeffFig1} for different values of the Hubble parameter. It is observed that the effective potential increases with increasing angular momentum, reaching a higher peak for larger $L$. The inclusion of $\alpha$ parameter shows the similar behavior. Furthermore, the graphical representation clearly shows the presence of a single maximum and the absence of any minimum in the effective potential profile. This indicates the existence of an unstable circular photon orbit at the location of the maximum.

\begin{figure}[htbp]
    \centering
    \includegraphics[width=0.42\textwidth]{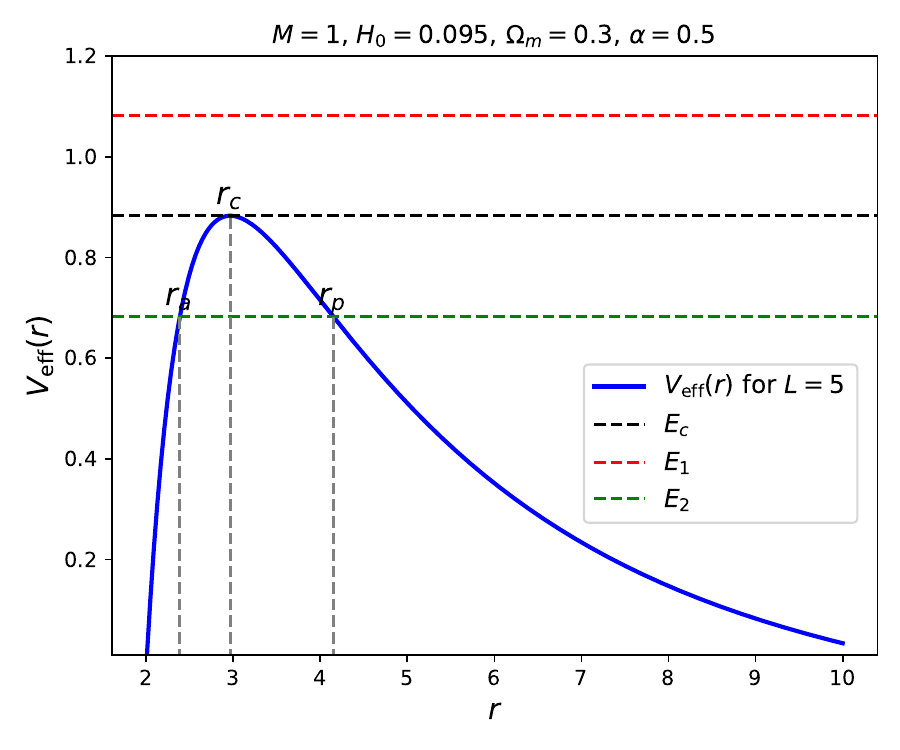}
    \caption{The behavior of the effective potential as a function of radial coordinate for different energy levels,highlighting the possible types of particle orbits.} \label{VeffOrbitsFig3}
\end{figure}

\noindent Fig.~\ref{VeffOrbitsFig3} illustrates the behavior of the effective potential $V_{\mathrm{eff}}(r)$ for photons along with the possible types of orbit, depending on the total energy $E$. We classify the cases as follows: 
\begin{itemize}
    \item \textbf{Case 1: $E = E_c$}; In this case, $E^2 - V_{\mathrm{eff}} = 0$ and $\dot{r} = 0$, leading to circular photon orbits. However, since the potential has a maximum at $r = r_c$, these are \emph{unstable circular orbits}, as shown in Fig.~\ref{VeffOrbitsFig3}. 
    \item \textbf{Case 2: $E = E_2$}; Here, $E^2 - V_{\mathrm{eff}} \geq 0$ only in the regions $r \geq r_a$ and $r \leq r_p$ (Fig.~\ref{VeffOrbitsFig3}). If photons start at infinity, they reach a minimum approach distance $r_P$ before being scattered back to infinity, resulting in gravitational deflection. Alternatively, if photons start at $r = r_a$, they cross the event horizon at $r = r_{h}$ and fall into the singularity.
    \item \textbf{Case 3: $E = E_1$}; In this case, $E^2 - V_{\mathrm{eff}} > 0$ for all $r$, implying that photons coming from infinity inevitably cross the event horizon at $r = r_{h}$ and fall into the singularity at $r = a$. To plot the possible types of photon orbits, we characterize the motion in terms of the impact parameter $b$.
\end{itemize}

\begin{figure}[htbp]
    \centering
    \includegraphics[width=0.48\textwidth]{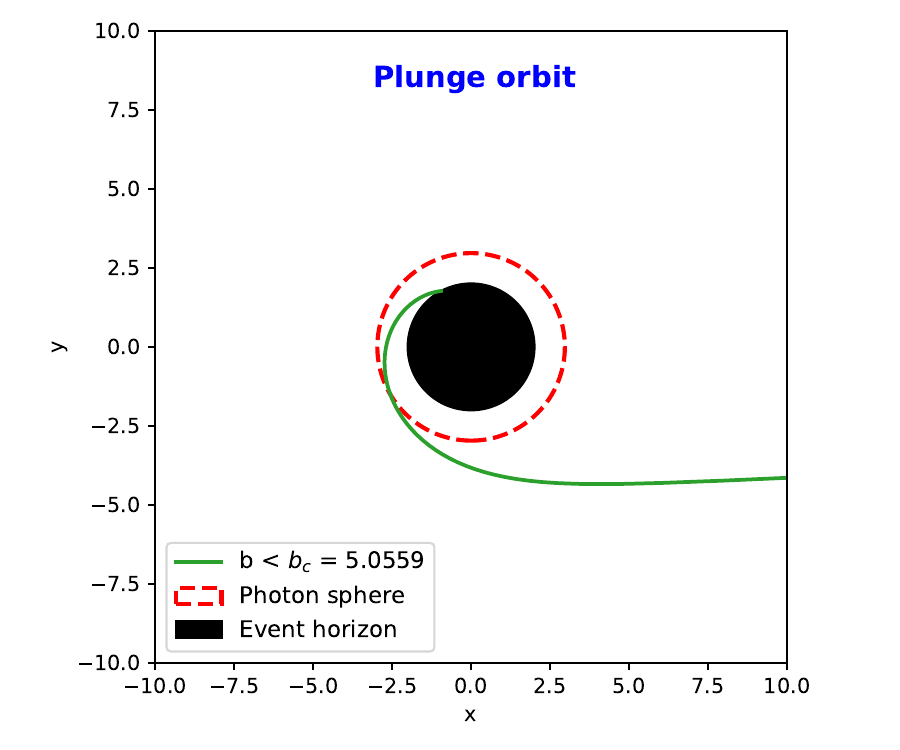}
    \includegraphics[width=0.48\textwidth]{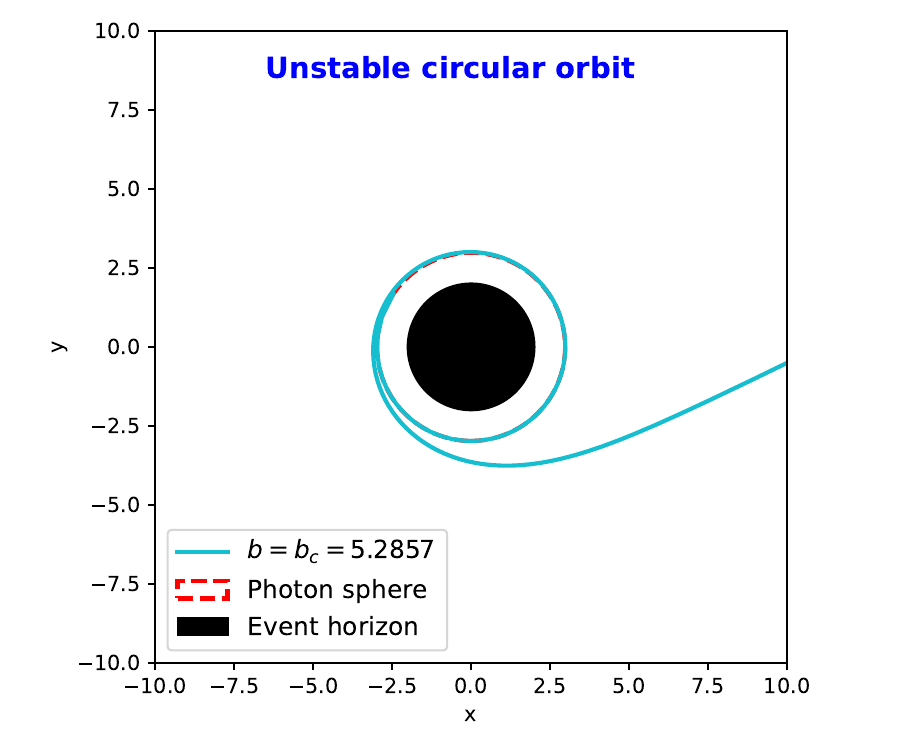}
    \includegraphics[width=0.48\textwidth]{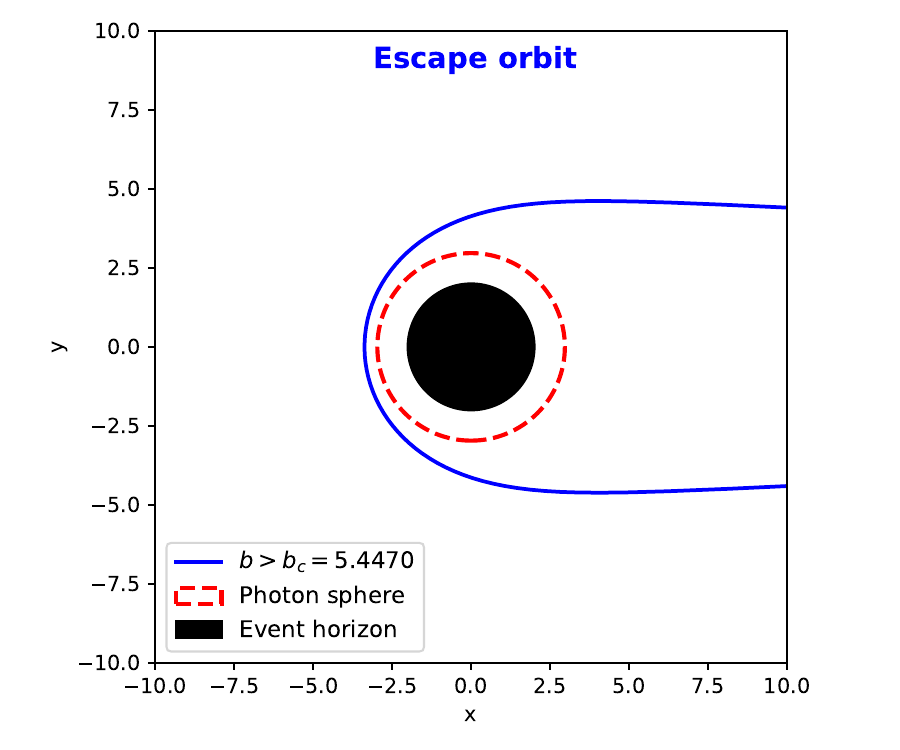}
    \caption{Classification of photon trajectories based on null geodesics: plunging orbits falling into the black hole, unstable circular photon orbit at the photon sphere, and escaping trajectories reaching infinity.} \label{VeffOrbitsFig3.1}
\end{figure}

\noindent Based on these cases, the possible type of orbits are shown in Fig.~\ref{VeffOrbitsFig3.1}. For $b < b_c$, light rays coming from infinity are gravitationally captured by the black hole, crossing the event horizon and plunging into the singularity (plunge-in orbits).  
For $b = b_c$, photons follow an unstable circular orbit at the photon sphere radius, where a small perturbation causes them to either escape to infinity or fall into the black hole.  
For $b > b_c$, photons are deflected by the gravitational field but ultimately escape to infinity (escape orbits).


\section{Deflection Angle}\label{Sec5}
In this section, we analyze the bending of light in the Schwarzschild-like spacetime inspired by cosmology using the Rindler–Ishak method. In contrast to the standard asymptotically flat case, the presence of cosmological modifications renders the usual definition of the deflection angle ambiguous. The Rindler–Ishak prescription provides a consistent framework by defining the deflection in terms of the local angle between the photon trajectory and the radial direction, making it suitable for non-asymptotically flat spacetimes.\\
For a static, spherically symmetric metric in the equatorial plane
\begin{equation}
 ds^{2} = -f(r)\,dt^{2} + \frac{dr^{2}}{f(r)} + r^{2} d\phi^{2},   
\end{equation}
the spatial line element on a constant--time slice is
\begin{equation}
 dl^{2} = \frac{dr^{2}}{f(r)} + r^{2} d\phi^{2}.   
\end{equation}
Let a light ray in this slice be parameterized by the trajectory \(r(\phi)\).
The angle \(\psi\) between the radial direction and the tangent to the photon
trajectory, as measured by a static observer at radius \(r\), is obtained from
the spatial metric by projecting the tangent vector onto the radial and
angular directions. A simple and commonly used expression (Rindler \& Ishak
2007 and subsequent works) is~\cite{Rindler:2007zz},
\begin{equation}\label{eq:tanpsi_definition}
    \tan\psi \;=\; \frac{\sqrt{g_{\phi\phi}}\,d\phi}{\sqrt{g_{rr}}\,dr}
    \;=\; r\sqrt{f(r)}\left|\frac{d\phi}{dr}\right|.
\end{equation}
We now express \(d\phi/dr\) in terms of the orbit equation. Using the inverse
radial coordinate \(u\equiv 1/r\), the null orbit equation for impact parameter
\(b\) is
\begin{equation}\label{eq:orbit_u}
    \left(\frac{du}{d\phi}\right)^2 \;=\; \frac{1}{b^2} - u^2 f(u).
\end{equation}
From \(u=1/r\) we have
\[
\frac{dr}{d\phi} = -\,\frac{1}{u^2}\frac{du}{d\phi}\quad\Longrightarrow\quad
\left|\frac{d\phi}{dr}\right|
= \frac{1}{r^{2}\sqrt{\frac{1}{b^{2}}-u^{2}f(u)}}.
\]
Substituting this into (\ref{eq:tanpsi_definition}) yields the compact form
\begin{equation}\label{eq:tanpsi_alt}
    \tan\psi(r) \;=\; \frac{b\sqrt{f(r)}}{\sqrt{\,r^{2}-b^{2}f(r)\,}}\,.
\end{equation}
Eq.~\ref{eq:tanpsi_alt} gives the
angle between the photon direction and the radial direction as measured by a
static observer located at radius \(r\). This expression is directly
applicable to your modified Schwarzschild lapse \(f(r)\) (or equivalently
\(f(u)\)) by substituting the metric function.

\paragraph{Measurable deflection between source and observer.}
Let the photon travel from a source at radius \(r_{s}\) to an observer at
radius \(r_{o}\). Integrate the orbit equation to obtain the coordinate
angular change~\cite{chandrasekhar1998mathematical},
\[
\Delta\phi \equiv \phi_{o}-\phi_{s} \;=\; \int_{r_{s}}^{r_{o}} \frac{d\phi}{dr}\,dr
        \;=\; \int_{r_{s}}^{r_{o}} \frac{dr}{r^{2}\sqrt{\dfrac{1}{b^{2}}-\dfrac{f(r)}{r^{2}}}}.
\]
The physically measurable bending between the ray and the radial directions
at source and observer is given by the difference of the local angles. One
useful and general definition (following the Rindler--Ishak method) is~\cite{Bhattacharya:2009rv},
\begin{equation}\label{eq:deflection_RI_general}
    \delta_{\rm meas} \;=\; \bigl[\psi(r_{o}) - \psi(r_{s})\bigr] \;+\; \Delta\phi,
\end{equation}
where \(\psi(r_{o})\) and \(\psi(r_{s})\) are computed from
Eq.~\ref{eq:tanpsi_alt} (taking the appropriate sign convention for
incoming or outgoing branches) and \(\Delta\phi\) is the coordinate angular
change computed from the orbit integral. Eq.~\ref{eq:deflection_RI_general}
is a coordinate-independent, observer-based measure of bending: it compares the
direction of the light ray relative to the local radial directions at the two
endpoints. If the source and the observer are at the same radius \(r_{s}=r_{o}=r\),
symmetric about the point of closest approach, the measurable bending reduces to $\delta_{\rm meas} \;=\; 2\psi(r) - \pi,$ which is the form often quoted in the Rindler--Ishak literature for a symmetric configuration.\\
To investigate the deflection angle of light in the modified Schwarzschild
metric, let us consider
\begin{align}
g(u) =\; & \left(1-\Omega_m^{\alpha+1}\right)^{\tfrac{1}{\alpha+1}} 
 + \frac{(r_c^3 u^3)^{\alpha+1}}
   {(\alpha+1)\left(1-\Omega_m^{\alpha+1}\right)^{\tfrac{\alpha}{\alpha+1}}} \nonumber\\[4pt]
& + \mathcal{O}\!\big((r_c^3 u^3)^{2(\alpha+1)}\big) .
\end{align}
and define the compact coefficients
\[
A \equiv \left(1-\Omega_m^{\alpha+1}\right)^{\frac{1}{\alpha+1}},\qquad
B \equiv \frac{1}{\alpha+1}\left(1-\Omega_m^{\alpha+1}\right)^{-\frac{\alpha}{\alpha+1}}.
\]
Using \(u=1/r\) and including the Schwarzschild mass term \(-2GM/r\) in the lapse, the metric function expanded for \(r\gg r_g\) reads
\begin{align}\label{eq:f_r_exp}
f(r) =\; & 1 - \frac{2GM}{r} 
- A\,H_0^2 r^2 
 - B\,H_0^2\,r_c^{3(\alpha+1)}\,r^{\,2-3(\alpha+1)} \nonumber\\[4pt]
& + \mathcal{O}\!\left(H_0^2 r_c^{6(\alpha+1)} 
     r^{\,2-6(\alpha+1)}\right) .
\end{align}
For a static observer at radius \(r\), the angle between the photon direction and the radial direction is given as
\[
\tan\psi(r)\;=\;\frac{b\sqrt{f(r)}}{\sqrt{\,r^{2}-b^{2}f(r)\,}}
\]
is expanded by writing \(f(r)=1-\delta(r)\) with
\[
\delta(r)=\frac{2GM}{r}+A H_0^2 r^2 + B H_0^2 r_c^{3(\alpha+1)} r^{\,2-3(\alpha+1)} + \cdots ,
\]
and assuming \(\delta\ll 1\) and \(s\equiv b/r\ll1\). Retaining terms up to linear order in \(\delta\) and up to \(s^3\) in geometric smallness, one finds
\begin{align}
    \tan\psi(r) &\simeq \; s \Bigg[ 1 - \frac{\delta(r)}{2} + \frac{s^{2}}{2} \Bigg] + \mathcal{O}(s^5,\; s \delta^2) .
\end{align}
Since angles are small we set \(\psi(r)\simeq \tan\psi(r)\). Thus
\begin{equation}\label{eq:psi_general}   
    \psi(r)\;\simeq\; \frac{b}{r} + \frac{b^{3}}{2r^{3}}
    - \frac{b}{2r}\,\delta(r) + \cdots \; 
\end{equation}
and substituting \(\delta(r)\) from Eq.~\ref{eq:f_r_exp} gives the explicit expansion
\begin{align}\label{eq:psi_explicit}
\psi(r) \simeq{}& 
\frac{b}{r} + \frac{b^{3}}{2r^{3}}
- \frac{b}{2r}\!\left(\frac{2GM}{r}\right)
- \frac{b}{2r}\!\left(A H_0^2 r^{2}\right) \nonumber\\[4pt]
& - \frac{b}{2r}\!\left(B H_0^2 r_c^{3(\alpha+1)} 
     r^{\,2-3(\alpha+1)}\right) + \cdots \nonumber\\[6pt]
=\;{}& \frac{b}{r} + \frac{b^{3}}{2r^{3}}
- \frac{b\,GM}{r^{2}}
- \frac{A}{2}\,H_0^2 b\, r \nonumber\\[4pt]
& - \frac{B}{2}\,H_0^2 b\, r_c^{3(\alpha+1)} 
     r^{\,1-3(\alpha+1)} + \cdots .
\end{align}

For the symmetric configuration \(r_s=r_o\) the Rindler--Ishak measurable deflection
is commonly taken (in the symmetric limit) as
\[
\delta_{\rm meas}\;=\;2\psi(r_o)\,,
\]
so combining Eq.~\ref{eq:psi_explicit} yields
\begin{align}\label{eq:Delta_meas_general}
\delta_{\rm meas} \simeq{}& 
\underbrace{\frac{4GM}{b}}_{\text{Schwarzschild}}
+ \underbrace{2\Big(\frac{b}{r_o}+\frac{b^{3}}{2r_o^{3}} - \frac{b\,GM}{r_o^{2}}\Big)}_{\text{finite--radius geometric corrections}} \nonumber\\[4pt]
& \underbrace{- A\,H_0^2\,b\,r_o
- B\,H_0^2\,b\,r_c^{3(\alpha+1)}\,r_o^{\,1-3(\alpha+1)}}_{\text{Cosmological corrections}} \nonumber\\[4pt]
& + \mathcal{O}\big(H_0^2 r_c^{6(\alpha+1)}\big).
\end{align}

In Eq.~\ref{eq:Delta_meas_general}, we have separated the familiar asymptotically flat leading term \(4GM/b\) (which arises from integrating the orbit) from the finite–radius geometric corrections (terms proportional to powers of \(b/r_o\) and \(GM/r_o^2\)), and from the explicit cosmological corrections proportional to \(H_0^2\). The terms proportional to \(H_0^2\) contain both a global contribution (appearing in the orbit integral and producing terms) and a local contribution coming from the local angle \(\psi(r_o)\); the above expression displays the local part explicitly as \(-A\,H_0^2 b r_o\) and the model-dependent decay as \(-B\,H_0^2 b\,r_c^{3(\alpha+1)} r_o^{1-3(\alpha+1)}\). 

\noindent Physically, these contributions signify that the presence of a nonzero Hubble constant modifies the bending of light even in the weak-field regime, producing a systematic deviation from the Schwarzschild prediction. While the Schwarzschild term scales solely with \( GM/b \), the additional cosmological terms introduce a scale-dependent correction, which grows with the observer’s distance \( r_o \) and depends sensitively on the model parameter \( \alpha \). This implies that, unlike the pure Schwarzschild black hole where lensing is entirely local, the bending angle in this 
scenario encodes imprints of the cosmic expansion, leading to potentially observable deviations from the standard Schwarzschild case in high-precision lensing measurements.

\begin{figure}[htbp] 
	\begin{center}
     {\includegraphics[width=0.45\textwidth,height=0.35\textheight]{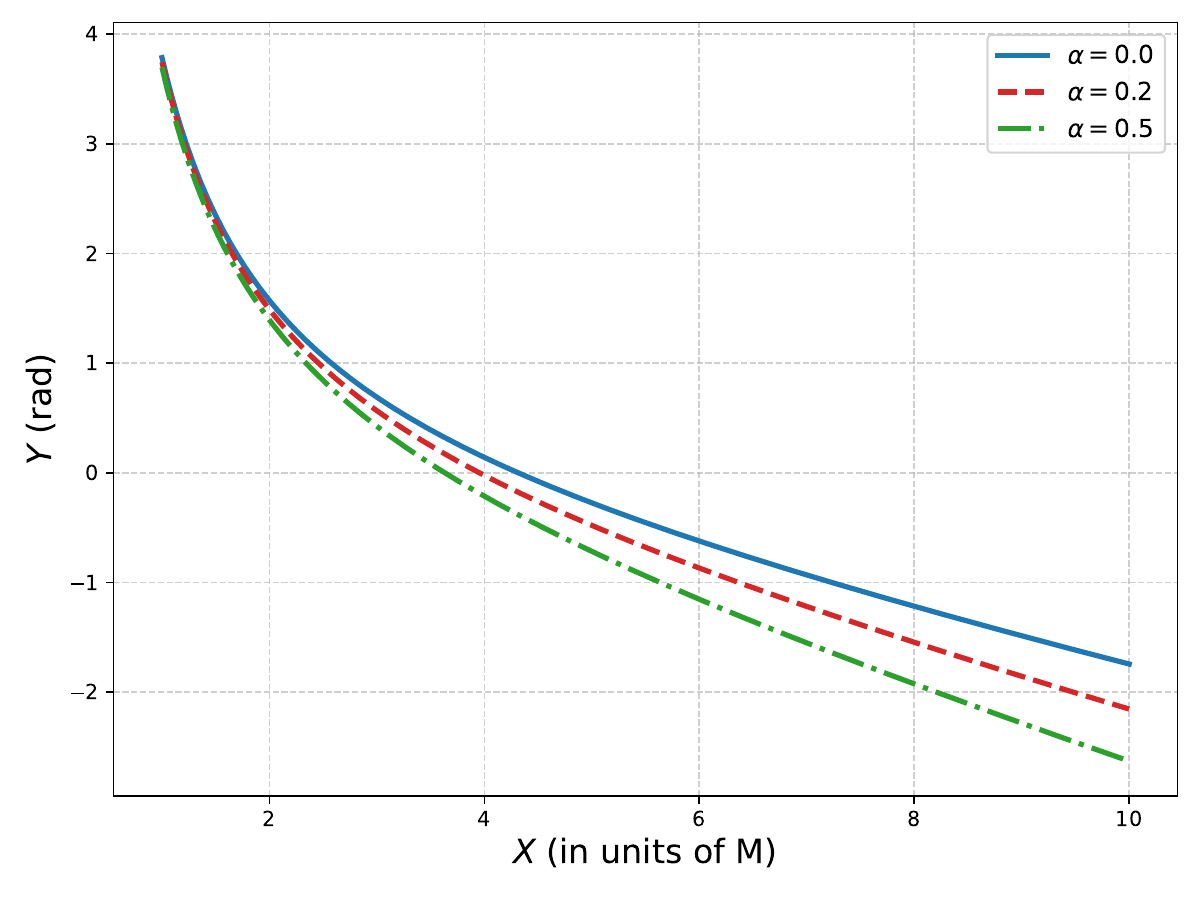}}   
	\end{center}
	\caption{The variation of the deflection angle (y-axis) as a function of the impact parameter (x-axis) for different values of $\alpha$. Here we consider, $H_{0}=0.095$, $\Omega_m=0.3$, and $r_{0}=20$.} \label{Fig4}
\end{figure}
\begin{figure}[htbp] 
	\begin{center}
     {\includegraphics[width=0.45\textwidth,height=0.35\textheight]{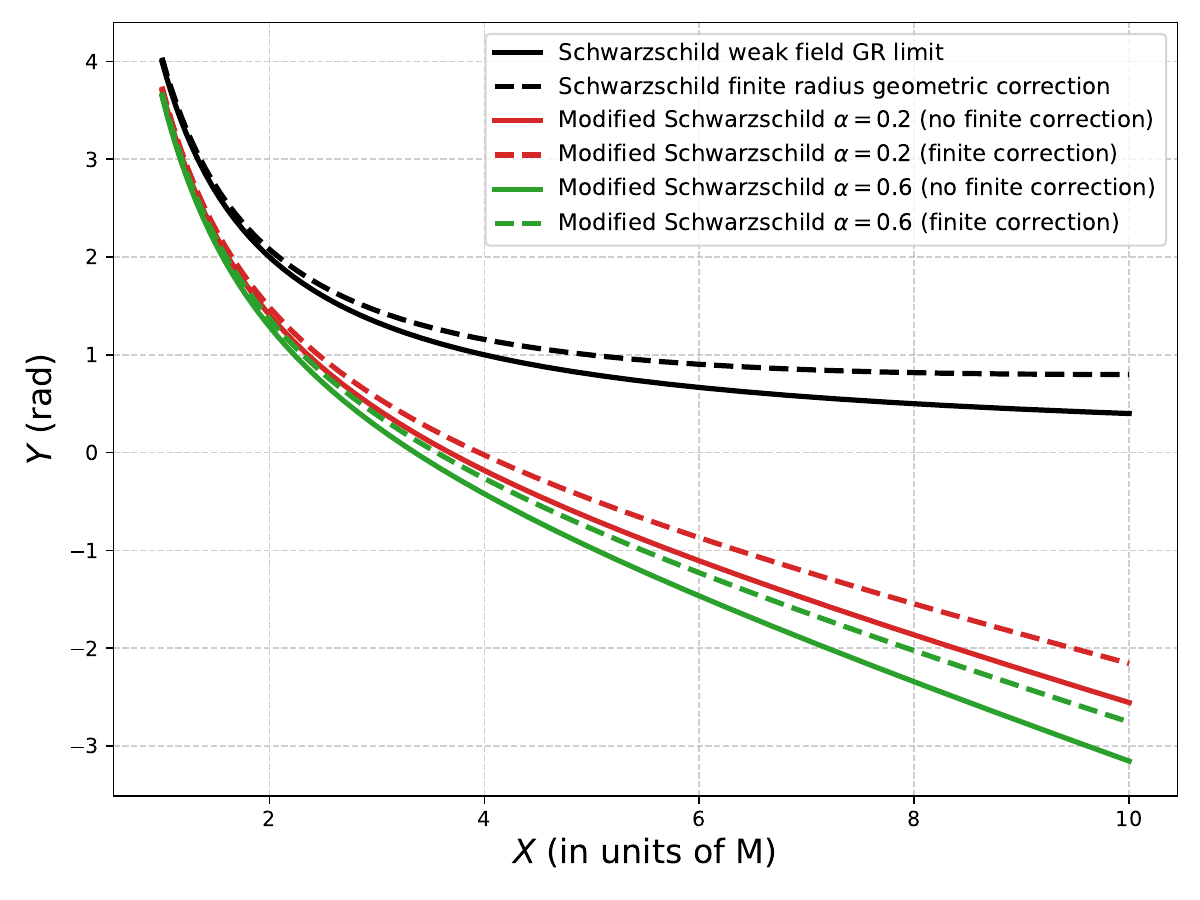}}   
	\end{center}
	\caption{Variation of the deflection angle as a function of the impact parameter for different values of $\alpha$, compared with the classical GR prediction for the Schwarzschild black hole in the weak-field limit. Here we consider, $H_{0}=0.095$, $\Omega_m=0.3$, and $r_{0}=20$.} \label{Fig5}
\end{figure}
The variation of the deflection angle with respect to the impact parameter $b$ for different values of the model parameter $\alpha$ exhibits a clear trend, as shown in Fig.~\ref{Fig4}. In all cases, the deflection angle decreases with increasing $b$, which is expected since photons passing farther from the black hole are less influenced by its gravity. The influence of $\alpha$ becomes apparent when comparing the curves: larger values of $\alpha$ systematically yield smaller deflection angles at a given $b$. This demonstrates that increasing $\alpha$ weakens the effective bending of light. Physically, this behavior arises from the modifications introduced by the model, which are governed by the cosmology-dependent parameters $H_0$ (the Hubble constant), $\Omega_m$ (the present-day matter density), and the crossover scale $r_c$. Together with $\alpha$, these parameters reduce the bending relative to the Schwarzschild case.\\
Fig.~\ref{Fig5} compares the deflection angle for the standard Schwarzschild black hole (SBH) and the modified SBH for $\alpha=0.2$ and $\alpha=0.6$, each plotted with and without finite-distance geometric corrections. The maximum deflection occurs first for SBH with finite-radius correction, followed by SBH without finite correction, then $\alpha=0.2$ with finite correction, $\alpha=0.2$ without finite correction, and similarly for $\alpha=0.6$. This ordering illustrates how both the finite-distance geometric effect and the modification parameter $\alpha$ systematically reduce the bending of light. Physically, it shows that finite-radius corrections enhance the deflection relative to the weak-field limit, while increasing $\alpha$ suppresses the bending, encoding the combined influence of local black hole geometry and cosmological modifications.\\
To emphasize observational relevance in a cosmological lensing context, we evaluate the case $\alpha = 1$ for a representative galaxy lens with mass $M = 10^{12}\,M_{\odot}$ and impact parameter $b = 10\ \mathrm{kpc}$, using $H_0 = 70\ \mathrm{km\,s^{-1}\,Mpc^{-1}}$, $\Omega_m = 0.315$, and an observer distance $r_0$. The classical general relativistic bending in this setup is $\hat{\delta}_{\rm GR} \approx 3.94834$ arcseconds. Incorporating finite-distance corrections from the observer at $r_0$, we solve for the characteristic scale $r_c$ that produces a micro-arcsecond-level correction, $\Delta \hat{\delta} \simeq 1 \times 10^{-6}$ arcseconds, giving $r_c \approx 1.30 \times 10^{17}\ \mathrm{m} \approx 4.2\ \mathrm{pc}$. At this scale, the fractional change becomes $\Delta \hat{\delta} / \hat{\delta}_{\rm GR} \sim 2.5 \times 10^{-7}$, with the higher-order $H_0^2 r_c^{6(\alpha+1)}$ term dominating the cosmology-induced correction. Physically, this indicates that cosmological expansion effects on light deflection are potentially testable only if the model allows a characteristic scale of order parsecs, in which case next-generation astrometric and lensing measurements with micro-arcsecond precision could provide meaningful constraints or possible detections. Such deviations, if present in nature, could be probed by current and future high-precision lensing measurements, such as those from the next generation EHT Collaboration~\cite{Johnson:2023ynn}, space-based missions like Euclid \cite{Euclid:2024yrr}, or upcoming surveys by the Vera C. Rubin Observatory \cite{Abrams:2023vzw}.
\section{Shadow observed by a static observer} \label{Sec6}
In this section, we briefly study the black hole shadow as observed by a static observer, following the approach of Perlick et al.~\cite{Perlick:2021aok}. Their framework allows computation of shadow size and shape in a spacetime influenced by cosmological expansion, including effects of a cosmological constant~\cite{Bisnovatyi-Kogan:2018vxl,Perlick:2018iye}. Using this method, we incorporate finite-distance corrections to the observer and evaluate how cosmological parameters and model modifications affect the observed shadow. This provides a physically meaningful connection between local black hole geometry and global cosmological effects.
Within the weak-field expansion of the lapse already used in the manuscript,
\begin{align}
f(r) = \; & 1 - \frac{2GM}{r} 
           - A\,H_{0}^{2} r^{2} 
           - B\,H_{0}^{2} r_{c}^{\,n} r^{\,2-n} \notag \\
         & + \, \mathcal{O}\!\left(H_{0}^{2} r_{c}^{2n} r^{\,2-2n}\right).
\end{align}
where for compactness we set
\[
A \equiv \Big(1-\Omega_m^{\alpha+1}\Big)^{\frac{1}{\alpha+1}},\qquad
B \equiv \frac{1}{\alpha+1}\Big(1-\Omega_m^{\alpha+1}\Big)^{-\frac{\alpha}{\alpha+1}},
\]
and \(n\equiv 3(\alpha+1)\). The term proportional to \(B\) encodes the
model-dependent higher-order (in \(r_c\)) contribution.

\paragraph{Photon sphere radius:}
Null circular (photon) orbits are obtained from the extremum condition of
the effective potential, equivalently
\begin{equation}\label{eq:photon_condition}
\frac{d}{dr}\!\left(\frac{f(r)}{r^{2}}\right) \;=\; 0
\quad\Longleftrightarrow\quad f'(r)\,r - 2f(r) \;=\; 0.
\end{equation}
Let \(r_{\rm ph}\) denote the solution. In the pure Schwarzschild limit
(\(H_0\!=\!0\)) one recovers \(r_{\rm ph}^{(0)} = 3GM\). Treating the
cosmological corrections perturbatively (small \(H_0^2\)), write
\(r_{\rm ph}=r_{\rm ph}^{(0)}+\Delta r\) with \(r_{\rm ph}^{(0)}=3GM\).
Expanding Eq.~\ref{eq:photon_condition} to first order in \(H_0^2\) yields
\begin{equation}\label{eq:rph_pert}
r_{\rm ph}
\;\simeq\; 3GM \left[1
    + \frac{n\,B\,H_{0}^{2}\,r_{c}^{\,n}\,(3GM)^{\,2-n}}{2}
    \;+\;\mathcal{O}(H_0^4)\right].
\end{equation}
Thus the photon-sphere radius is shifted from \(3GM\) by a small \(H_0^2\)
dependent correction; the sign and magnitude are controlled by \(nB\)
(and hence by \(\alpha\) and \(\Omega_m\)).

\paragraph{Critical impact parameter:}
The critical impact parameter \(b_{\rm crit}\) (separating plunging and
scattering null geodesics) is given in terms of the photon-sphere radius by
\begin{equation}\label{eq:bc_general}
b_{\rm crit} \;=\; \frac{r_{\rm ph}}{\sqrt{f(r_{\rm ph})}}.
\end{equation}
To leading (Schwarzschild) order \(f(r_{\rm ph}^{(0)})=1-2GM/(3GM)=1/3\), so
\(b_{\rm crit}^{(0)}=3GM/\sqrt{1/3}=3\sqrt{3}\,GM\). Including first-order
\(H_0^2\) corrections one may write
\begin{equation}\label{eq:bc_pert}
b_{\rm crit} \simeq 3\sqrt{3}\,GM
\left[1 + \frac{\Delta r}{3GM} - \frac{1}{2}\frac{\delta f(r_{\rm ph}^{(0)})}{f(r_{\rm ph}^{(0)})}
    + \mathcal{O}(H_0^4)\right],
\end{equation}
where \(\Delta r\) is the photon-sphere shift from Eq.~\ref{eq:rph_pert} and
\(\delta f(r_{\rm ph}^{(0)})\) denotes the \(H_0^2\)-piece of \(f\) evaluated
at \(r=3GM\),
\[
\delta f(3GM) = -A\,H_0^2 (3GM)^2 - B\,H_0^2 r_c^{\,n}(3GM)^{\,2-n}.
\]
Eqs.~\ref{eq:bc_general}--\eqref{eq:bc_pert} provide a controlled
perturbative approximation for \(b_{\rm crit}\) in terms of the model parameters of modified spacetime.

\paragraph{Angular radius as seen by a static observer:}
A static observer located at radius \(r_o\) measures the angular radius
\(\theta_{\rm sh}\) of the shadow through the local geometry. The angular radius of black hole shadow for a static observer located at a large but finite distance can be determined using the exact expression~\cite{Synge:1966okc},
\begin{equation}\label{eq:sin_theta_exact}
\sin\theta_{\rm sh} \;=\; \frac{b_{\rm crit}\,\sqrt{f(r_o)}}{r_o}\;,
\end{equation}
which is equivalent to the tangent formula \(\tan\psi = b\sqrt{f}/\sqrt{r^2-b^2 f}\)
used earlier (both expressions are algebraically related and give the same
numerical result). For small angles one may use \(\theta_{\rm sh}\simeq
b_{\rm crit}\sqrt{f(r_o)}/r_o\).

Substituting the perturbative expression for \(b_{\rm crit}\) and expanding
to leading nontrivial order yields
 \begin{align}
\theta_{\rm sh} &\;\simeq\; \frac{3\sqrt{3}\,GM}{r_o}\,\sqrt{f(r_o)} \,  \nonumber\\[4pt]
& \times \Bigg[ 1 + \frac{\Delta r}{3GM} 
- \frac{1}{2}\frac{\delta f(r_{\rm ph}^{(0)})}{f(r_{\rm ph}^{(0)})} 
+ \mathcal{O}\big(H_0^4, (GM/r_o)^2\big) \Bigg]. \label{eq:theta_pert} 
\end{align}   
\begin{figure}[htbp]
    \centering
    \includegraphics[width=0.5\textwidth,height=0.35\textheight]{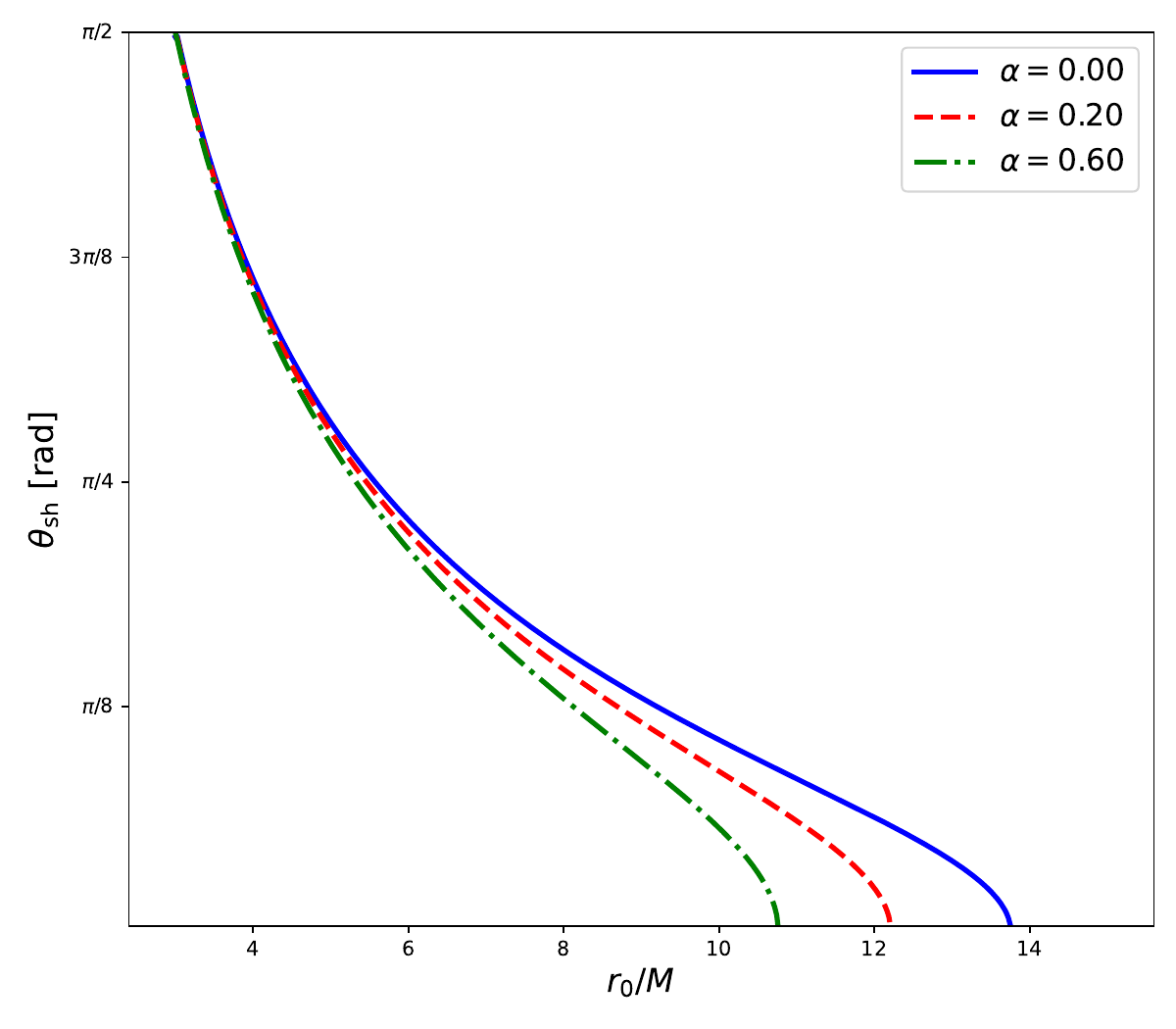}
    \caption{Angular radius of the shadow plotted against the observer position. Here we consider, $H_{0}=0.095$, and $\Omega_m=0.3$. }
    \label{fig:statshadow}
\end{figure}
We have plotted the variation of the angular shadow radius of the black hole as a function of the observer’s position in Fig.~\ref{fig:statshadow}. It is observed that the shadow radius attains its maximum value, approaching $\pi/2$, near the black hole horizon. As the observer moves farther away, the angular shadow radius decreases monotonically and eventually tends to zero at the cosmological horizon. This behavior is physically intuitive: close to the event horizon, the black hole subtends the largest possible angular size in the observer’s sky, while at large distances the shadow gradually shrinks and vanishes as the spacetime approaches asymptotic cosmological scales. Furthermore, the presence of the model parameter reduces the angular shadow radius for a given observer distance, indicating that the additional effects of the parameter effectively diminish the apparent size of the shadow. This highlights how modifications to the black hole spacetime imprint themselves on observable quantities such as the angular size of the shadow. Our results are in excellent qualitative agreement with the study of black hole shadows at local and cosmological distances by Bisnovatyi-Kogan et al.~\cite{Bisnovatyi-Kogan:2019wdd}, where the shadow size is shown to be maximal near the event horizon and gradually decreases to vanish at the cosmological horizon.

\section{Conclusion} \label{Sec7}
In this paper, we studied cosmology-inspired modifications to Schwarzschild-like black hole spacetime and their effects on photon trajectories, gravitational lensing, and shadow observables. The main findings are summarized as follows:
\begin{itemize}
    \item Cosmology-inspired modifications were investigated through the lapse function \(f(r)\). The deviation parameter \(\alpha\) does not eliminate or merge horizons; across the considered range, the spacetime always admits both an inner black hole horizon and an outer cosmological horizon, preserving a non-extremal black hole structure. Thus, \(\alpha\) only shifts the radial positions of the horizons without changing their qualitative nature.  

    \item The effective potential for massless particles exhibits a single maximum corresponding to the unstable circular photon orbit. Increasing angular momentum shifts this maximum to higher potential values. Photon trajectories are classified into plunge-in, unstable circular, and escape orbits depending on the impact parameter relative to the critical value.  

    \item The deflection angle decreases with increasing impact parameter \(b\), as expected since photons passing farther from the black hole experience weaker gravity. Larger values of the deviation parameter \(\alpha\) systematically yield smaller deflection angles, demonstrating that cosmology-inspired modifications suppress light bending relative to the Schwarzschild case. Finite-distance geometric corrections, computed via the Rindler–Ishak method, enhance the bending compared to the asymptotic treatment, but this effect is progressively reduced as \(\alpha\) increases. Physically, this trend reflects the combined influence of local black hole geometry and cosmological parameters \((H_{0}, \Omega_{m}, r_{c}, \alpha)\) on photon trajectories.

    \item For a representative galactic lens with parameters \(\alpha = 1\), \(M = 10^{12}\,M_{\odot}\), and \(b = 10\,\mathrm{kpc}\), the classical deflection angle is \(\hat{\delta}_{\rm GR} \approx 3.95\) arcseconds.\\
    Incorporating cosmological corrections with \(H_{0} = 70\,\mathrm{km\,s^{-1}\,Mpc^{-1}}\) and \(\Omega_{m} = 0.315\), a characteristic scale \(r_{c} \approx 4.2\) pc produces a micro-arcsecond deviation, dominated by the higher-order $H_0^2 r_c^{6(\alpha+1)}$ term.  

    \item The angular radius of the black hole shadow, as seen by a static observer, is largest near the event horizon and decreases steadily as the observer moves outward, eventually vanishing at the cosmological horizon. This behavior is physically intuitive, since the black hole subtends its greatest angular size when viewed close to the horizon, while at large distances the shadow shrinks due to cosmological effects. The presence of the model parameter further reduces the angular size for a given observer distance, showing that cosmology-inspired modifications diminish the apparent shadow. These results are consistent with earlier studies, confirming that the shadow is maximal near the horizon and disappears at the cosmological boundary.
     
\end{itemize}
The overall results of this work demonstrate that cosmo-logy-inspired modifications have a notable impact on both gravitational light deflection and black hole shadow characteristics. These effects underscore the observational relevance of our study, indicating that high precision measurements of photon trajectories and shadow morphology can provide meaningful constraints on deviations from the standard Schwarzschild geometry and offer a probe of cosmological influences in strong-field gravitational regimes.
\section*{Acknowledgments}
The author SK acknowledges the contribution of the COST Action CA21136 – “Addressing observational tensions in cosmology with systematics and fundamental physics (CosmoVerse)”. HN would like to acknowledge the financial support provided by the Anusandhan National Research Foundation (ANRF), New Delhi, thr-ough the grant number CRG/2023/008980. This work has also been supported by the project i-COOPB23096, funded by CSIC.

\appendix
\section{Dimensionless Hubble Parameter in Geometric Units}

In this work, we adopt geometric units in which \(G = c = 1\) and all lengths are measured in units of the black hole mass \(M\). In these units, the Hubble constant must be rendered dimensionless to be compatible with the metric functions.  

The physical Hubble constant, \(H_0^{\rm phys}\), has dimensions of inverse time (s\(^{-1}\)). To convert it to dimensionless form, we use the natural time scale of the black hole,
\begin{equation}
t_{\rm BH} = \frac{G M}{c^3}.    
\end{equation}
The dimensionless Hubble parameter is then defined as
\begin{equation}
 H_0 = H_0^{\rm phys} \, t_{\rm BH} = H_0^{\rm phys} \, \frac{G M}{c^3}.   
\end{equation}
For example, for a black hole of mass \(M\) and a physical Hubble constant \(H_0^{\rm phys} \simeq 70~{\rm km~s^{-1}~Mpc^{-1}}\), the corresponding dimensionless value in geometric units is
\begin{equation}
 H_0 \simeq 2.27 \times 10^{-18} \, {\rm s^{-1}} \times \frac{G M}{c^3}.   
\end{equation}
Using this dimensionless form ensures that all terms in the Schwarzschild-like metric and related equations are consistent and free of explicit physical units.  

\bibliographystyle{unsrt}
\bibliography{main}

\begin{thebibliography}{10}

\bibitem{SupernovaSearchTeam:1998fmf}
Adam~G. Riess et~al.
\newblock {Observational evidence from supernovae for an accelerating universe and a cosmological constant}.
\newblock {\em Astron. J.}, 116:1009--1038, 1998.

\bibitem{Chevallier:2000qy}
Michel Chevallier and David Polarski.
\newblock {Accelerating universes with scaling dark matter}.
\newblock {\em Int. J. Mod. Phys. D}, 10:213--224, 2001.

\bibitem{DESI:2025wyn}
Gan Gu et~al.
\newblock {Dynamical Dark Energy in light of the DESI DR2 Baryonic Acoustic Oscillations Measurements}.
\newblock {\em FERMILAB-PUB-25-0235-PPD}, 4 2025.

\bibitem{Chudaykin:2024gol}
Anton Chudaykin and Martin Kunz.
\newblock {Modified gravity interpretation of the evolving dark energy in light of DESI data}.
\newblock {\em Phys. Rev. D}, 110(12):123524, 2024.

\bibitem{Padmanabhan:2004av}
T.~Padmanabhan.
\newblock {Dark energy: The Cosmological challenge of the millennium}.
\newblock {\em Curr. Sci.}, 88:1057, 2005.

\bibitem{Straumann:2006tv}
Norbert Straumann.
\newblock {Dark Energy: Recent Developments}.
\newblock {\em Mod. Phys. Lett. A}, 21:1083--1098, 2006.

\bibitem{dvali20004d}
Gia Dvali, Gregory Gabadadze, and Massimo Porrati.
\newblock 4d gravity on a brane in 5d minkowski space.
\newblock {\em Phy. Lett. B}, 485(1-3):208--214, 2000.

\bibitem{freese2002cardassian}
Katherine Freese and Matthew Lewis.
\newblock Cardassian expansion: a model in which the universe is flat, matter dominated, and accelerating.
\newblock {\em Phy. Lett. B}, 540(1-2):1--8, 2002.

\bibitem{dvali:2003rk}
Gia Dvali and Michael~S. Turner.
\newblock {Dark Energy as a Modification of the Friedmann Equation}.
\newblock {\em FERMILAB-PUB-03-040-A}, 1 2003.

\bibitem{lue2004differentiating}
Arthur Lue, Roman Scoccimarro, and Glenn Starkman.
\newblock Differentiating between modified gravity and dark energy.
\newblock {\em Phys. Rev. D}, 69(4):044005, 2004.

\bibitem{barreiro2004generalized}
T~Barreiro and Anjan~Ananda Sen.
\newblock Generalized chaplygin gas in a modified gravity approach.
\newblock {\em Phys. Rev. D}, 70(12):124013, 2004.

\bibitem{hartle2021gravity}
James~B Hartle.
\newblock {\em {Gravity: an introduction to Einstein's general relativity}}.
\newblock Cambridge University Press, 2021.

\bibitem{poisson2004relativist}
Eric Poisson.
\newblock {\em {A relativist's toolkit: the mathematics of black-hole mechanics}}.
\newblock Cambridge university press, 2004.

\bibitem{wald2010general}
Robert~M Wald.
\newblock {\em General relativity}.
\newblock University of Chicago press, 2010.

\bibitem{chandrasekhar1998mathematical}
Subrahmanyan Chandrasekhar.
\newblock {\em The mathematical theory of black holes}, volume~69.
\newblock Oxford university press, 1998.

\bibitem{Breton:2002td}
N.~Breton.
\newblock {Geodesic structure of the Born-Infeld black hole}.
\newblock {\em Class. Quant. Grav.}, 19:601--612, 2002.

\bibitem{hackmann2008geodesic}
Eva Hackmann and Claus Lammerzahl.
\newblock {Geodesic equation in Schwarzschild- (anti-) de Sitter space-times: Analytical solutions and applications}.
\newblock {\em Phys. Rev. D}, 78:024035, 2008.

\bibitem{Grunau:2010gd}
Saskia Grunau and Valeria Kagramanova.
\newblock {Geodesics of electrically and magnetically charged test particles in the Reissner-Nordstr{\"o}m space-time: analytical solutions}.
\newblock {\em Phys. Rev. D}, 83:044009, 2011.

\bibitem{hackmann2010geodesic}
Eva Hackmann.
\newblock {\em Geodesic equations in black hole space-times with cosmological constant}.
\newblock PhD thesis, Universit{\"a}t Bremen, 2010.

\bibitem{Abbas:2014oua}
G.~Abbas and U.~Sabiullah.
\newblock {Geodesic Study of Regular Hayward Black Hole}.
\newblock {\em Astrophys. Space Sci.}, 352:769--774, 2014.

\bibitem{uniyal2015geodesic}
Rashmi Uniyal, N~Chandrachani~Devi, Hemwati Nandan, and KD~Purohit.
\newblock Geodesic motion in schwarzschild spacetime surrounded by quintessence.
\newblock {\em Gen. Rel. Grav.}, 47(2):16, 2015.

\bibitem{Soroushfar:2016yea}
Saheb Soroushfar, Reza Saffari, and Ehsan Sahami.
\newblock {Geodesic equations in the static and rotating dilaton black holes: Analytical solutions and applications}.
\newblock {\em Phys. Rev. D}, 94(2):024010, 2016.

\bibitem{Kala:2021ppi}
Shubham Kala, Hemwati Nandan, Prateek Sharma, and Maye Elmardi.
\newblock {Geodesics and bending of light around a BTZ black hole surrounded by quintessential matter}.
\newblock {\em Mod. Phys. Lett. A}, 36(31):2150224, 2021.

\bibitem{Battista:2022krl}
Emmanuele Battista and Giampiero Esposito.
\newblock {Geodesic motion in Euclidean Schwarzschild geometry}.
\newblock {\em Eur. Phys. J. C}, 82(12):1088, 2022.

\bibitem{Capozziello:2025wwl}
Salvatore Capozziello, Emmanuele Battista, and Silvia De~Bianchi.
\newblock {Null geodesics, causal structure, and matter accretion in Lorentzian-Euclidean black holes}.
\newblock {\em Phys. Rev. D}, 112(4):044009, 2025.

\bibitem{Virbhadra:1999nm}
K.~S. Virbhadra and George F.~R. Ellis.
\newblock {Schwarzschild black hole lensing}.
\newblock {\em Phys. Rev. D}, 62:084003, 2000.

\bibitem{Bozza:2001xd}
V.~Bozza, S.~Capozziello, G.~Iovane, and G.~Scarpetta.
\newblock {Strong field limit of black hole gravitational lensing}.
\newblock {\em Gen. Rel. Grav.}, 33:1535--1548, 2001.

\bibitem{Iyer:2006cn}
Savitri~V. Iyer and Arlie~O. Petters.
\newblock {Light's bending angle due to black holes: From the photon sphere to infinity}.
\newblock {\em Gen. Rel. Grav.}, 39:1563--1582, 2007.

\bibitem{Cunha:2018acu}
Pedro V.~P. Cunha and Carlos A.~R. Herdeiro.
\newblock {Shadows and strong gravitational lensing: a brief review}.
\newblock {\em Gen. Rel. Grav.}, 50(4):42, 2018.

\bibitem{Kala:2020prt}
Shubham Kala, Saurabh, Hemwati Nandan, and Prateek Sharma.
\newblock {Deflection of light and shadow cast by a dual-charged stringy black hole}.
\newblock {\em Int. J. Mod. Phys. A}, 35(28):2050177, 2020.

\bibitem{Kala:2020viz}
Shubham Kala, Hemwati Nandan, and Prateek Sharma.
\newblock {Deflection of Light Around a Rotating BTZ Black Hole}.
\newblock {\em Mod. Phys. Lett. A}, 35(39):2050323, 2020.

\bibitem{Kala:2022uog}
Shubham Kala, Hemwati Nandan, and Prateek Sharma.
\newblock {Shadow and weak gravitational lensing of a rotating regular black hole in a non-minimally coupled Einstein-Yang-Mills theory in the presence of plasma}.
\newblock {\em Eur. Phys. J. Plus}, 137(4):457, 2022.

\bibitem{Kala:2024fvg}
Shubham Kala, Hemwati Nandan, Amare Abebe, and Saswati Roy.
\newblock {Gravitational lensing around a dual-charged stringy black hole in plasma background}.
\newblock {\em Eur. Phys. J. C}, 84(10):1089, 2024.

\bibitem{Kala:2025fld}
Shubham Kala and Jaswinder Singh.
\newblock {Gravitational lensing and shadow around a non-minimally coupled Horndeski black hole in plasma medium (Accepted in Eur. Phys. J. C)}.
\newblock {\em arxiv:2507.17280}, 7 2025.

\bibitem{Roy:2025hdw}
Saswati Roy, Shubham Kala, Atanu Singha, Hemwati Nandan, and Asoke~K. Sen.
\newblock {Deflection of light due to Kerr Sen black hole in heterotic string theory using material medium approach}.
\newblock {\em Eur. Phys. J. C}, 85(7):772, 2025.

\bibitem{Roy:2025qmx}
Saswati Roy, Shubham Kala, Prasanjit Ghosh, Hemwati Nandan, and Asoke~K. Sen.
\newblock {Non-equatorial deflection of light due to Kerr{\textendash}Newman black hole: a material medium approach}.
\newblock {\em Eur. Phys. J. C}, 85(8):925, 2025.

\bibitem{Kala:2025iri}
Shubham Kala.
\newblock {Propagation of massless particles around a BTZ-ModMax black hole}.
\newblock {\em arXiv:2501.15999}, 2025.

\bibitem{Kukreti:2025rzn}
Shrishti Kukreti, Shubham Kala, Hemwati Nandan, Faizuddin Ahmed, and Saswati Roy.
\newblock {Equatorial light bending around a Hairy Kiselev Black Hole (Accepted in Nuclear Physics B)}.
\newblock {\em arxiv:2507.15298}, 7 2025.

\bibitem{Feleppa:2024kio}
Fabiano Feleppa, Valerio Bozza, and Oleg~Yu. Tsupko.
\newblock {Strong deflection of massive particles in spherically symmetric spacetimes}.
\newblock {\em Phys. Rev. D}, 111(4):044018, 2025.

\bibitem{Feleppa:2024vdk}
Fabiano Feleppa, Valerio Bozza, and Oleg~Yu. Tsupko.
\newblock {Strong deflection limit analysis of black hole lensing in inhomogeneous plasma}.
\newblock {\em Phys. Rev. D}, 110(6):064031, 2024.

\bibitem{Ahmed:2023dvc}
Faizuddin Ahmed.
\newblock {Geodesics motion of test particles around Schwarzschild-Klinkhamer wormhole with topological defects and gravitational lensing}.
\newblock {\em JCAP}, 11:010, 2023.

\bibitem{Ahmed:2024fye}
Faizuddin Ahmed.
\newblock {Gravitational lensing in holonomy corrected spherically symmetric black holes with phantom global monopoles}.
\newblock {\em Int. J. Geom. Meth. Mod. Phys.}, 22(05):2450336, 2025.

\bibitem{Pantig:2024lpg}
Reggie~C. Pantig, Shubham Kala, Ali {\"O}vg{\"u}n, and Nikko John Leo~S. Lobos.
\newblock {Testing black holes with cosmological constant in Einstein-bumblebee gravity through the black hole shadow using EHT data and deflection angle}.
\newblock {\em Int. J. Geom. Meth. Mod. Phys.}, 10 2024.

\bibitem{Vishvakarma:2024icz}
Bijendra~Kumar Vishvakarma, Shubham Kala, and Sanjay Siwach.
\newblock {Strong gravitational lensing by Bardeen black hole in cloud of strings}.
\newblock {\em Annals Phys.}, 475:169957, 2025.

\bibitem{ahmed2025gravitational}
Faizuddin Ahmed and Shubham Kala.
\newblock Gravitational lensing and topological photon sphere of holonomy corrected schwarzschild black hole with a cloud of strings.
\newblock {\em arXiv:2509.07686}, 2025.

\bibitem{Waseem:2025yib}
Hira Waseem, Nikko John Leo~S. Lobos, Ali {\"O}vg{\"u}n, and Reggie~C. Pantig.
\newblock {Analyzing deflection angles and photon sphere dynamics of magnetically charged black holes in nonlinear electrodynamic}.
\newblock {\em Eur. Phys. J. C}, 85(6):629, 2025.

\bibitem{Lambiase:2024uzy}
Gaetano Lambiase, Reggie~C. Pantig, and Ali {\"O}vg{\"u}n.
\newblock {Weak field deflection angle and analytical parameter estimation of the Lorentz-violating Bumblebee parameter through the black hole shadow using EHT data}.
\newblock {\em EPL}, 148(4):49001, 2024.

\bibitem{Turakhonov:2025ojy}
Ziyodulla Turakhonov, Farruh Atamurotov, Sushant~G. Ghosh, and Ahmadjon Abdujabbarov.
\newblock {Probing effects of plasma on shadow and weak gravitational lensing by regular black holes in asymptotically safe gravity}.
\newblock {\em Phys. Dark Univ.}, 48:101880, 2025.

\bibitem{Blandford:1991xc}
R.~D. Blandford and R.~Narayan.
\newblock {Cosmological applications of gravitational lensing}.
\newblock {\em Ann. Rev. Astron. Astrophys.}, 30:311--358, 1992.

\bibitem{Ratra:1992ut}
Bharat Ratra and Alice Quillen.
\newblock {Gravitational lensing effects in a time variable cosmological 'constant' cosmology}.
\newblock {\em Mon. Not. Roy. Astron. Soc.}, 259:738, 1992.

\bibitem{Bacon:2000sy}
David~J. Bacon, Alexandre~R. Refregier, and Richard~S. Ellis.
\newblock {Detection of weak gravitational lensing by large-scale structure}.
\newblock {\em Mon. Not. Roy. Astron. Soc.}, 318:625, 2000.

\bibitem{Hoekstra:2008db}
Henk Hoekstra and Bhuvnesh Jain.
\newblock {Weak Gravitational Lensing and its Cosmological Applications}.
\newblock {\em Ann. Rev. Nucl. Part. Sci.}, 58:99--123, 2008.

\bibitem{Aghili:2014aga}
Mir~Emad Aghili, Brett Bolen, and Luca Bombelli.
\newblock {Effect of accelerated global expansion on the bending of light}.
\newblock {\em Gen. Rel. Grav.}, 49(1):10, 2017.

\bibitem{Islam:2021ful}
Shafqat~Ul Islam, Jitendra Kumar, and Sushant~G. Ghosh.
\newblock {Strong gravitational lensing by rotating Simpson-Visser black holes}.
\newblock {\em JCAP}, 10:013, 2021.

\bibitem{Kumar:2021cyl}
Jitendra Kumar, Shafqat~Ul Islam, and Sushant~G. Ghosh.
\newblock {Investigating strong gravitational lensing effects by supermassive black holes with Horndeski gravity}.
\newblock {\em Eur. Phys. J. C}, 82(5):443, 2022.

\bibitem{Wang:2024iwt}
Yiyang Wang, Amnish Vachher, Qiang Wu, Tao Zhu, and Sushant~G. Ghosh.
\newblock {Strong gravitational lensing by static black holes in effective quantum gravity}.
\newblock {\em Eur. Phys. J. C}, 85(3):302, 2025.

\bibitem{Synge:1966okc}
J.~L. Synge.
\newblock {The Escape of Photons from Gravitationally Intense Stars}.
\newblock {\em Mon. Not. Roy. Astron. Soc.}, 131(3):463--466, 1966.

\bibitem{Perlick:2018iye}
Volker Perlick, Oleg~Yu. Tsupko, and Gennady~S. Bisnovatyi-Kogan.
\newblock {Black hole shadow in an expanding universe with a cosmological constant}.
\newblock {\em Phys. Rev. D}, 97(10):104062, 2018.

\bibitem{Bisnovatyi-Kogan:2018vxl}
Gennady~S. Bisnovatyi-Kogan and Oleg~Yu. Tsupko.
\newblock {Shadow of a black hole at cosmological distances}.
\newblock {\em Phys. Rev. D}, 98(8):084020, 2018.

\bibitem{Bisnovatyi-Kogan:2019wdd}
G.~S. Bisnovatyi-Kogan, O.~Yu. Tsupko, and V.~Perlick.
\newblock {Shadow of a black hole at local and cosmological distances}.
\newblock {\em PoS}, MULTIF2019:009, 2019.

\bibitem{EventHorizonTelescope:2019dse}
Kazunori Akiyama et~al.
\newblock {First M87 Event Horizon Telescope Results. I. The Shadow of the Supermassive Black Hole}.
\newblock {\em Astrophys. J. Lett.}, 875:L1, 2019.

\bibitem{Li:2020drn}
Peng-Cheng Li, Minyong Guo, and Bin Chen.
\newblock {Shadow of a Spinning Black Hole in an Expanding Universe}.
\newblock {\em Phys. Rev. D}, 101(8):084041, 2020.

\bibitem{Cotaescu:2020kcr}
Ion~I. Cotaescu.
\newblock {Light from Schwarzschild Black Holes in de Sitter expanding universe}.
\newblock {\em Eur. Phys. J. C}, 81(1):32, 2021.

\bibitem{Cuzinatto:2023kbo}
R.~R. Cuzinatto, C.~A.~M. de~Melo, and Juliano C.~S. Neves.
\newblock {Shadows of black holes at cosmological distances in the co-varying physical couplings framework}.
\newblock {\em Mon. Not. Roy. Astron. Soc.}, 526(3):3987--3993, 2023.

\bibitem{Rindler:2007zz}
Wolfgang Rindler and Mustapha Ishak.
\newblock {Contribution of the cosmological constant to the relativistic bending of light revisited}.
\newblock {\em Phys. Rev. D}, 76:043006, 2007.

\bibitem{Bhattacharya:2009rv}
Amrita Bhattacharya, Alexey Panchenko, Massimo Scalia, Carlo Cattani, and Kamal~K. Nandi.
\newblock {Light bending in the galactic halo by Rindler-Ishak method}.
\newblock {\em JCAP}, 09:004, 2010.

\bibitem{Johnson:2023ynn}
Michael~D. Johnson et~al.
\newblock {Key Science Goals for the Next-Generation Event Horizon Telescope}.
\newblock {\em Galaxies}, 11(3):61, 2023.

\bibitem{Euclid:2024yrr}
Y.~Mellier et~al.
\newblock {Euclid. I. Overview of the Euclid mission}.
\newblock {\em Astron. Astrophys.}, 697:A1, 2025.

\bibitem{Abrams:2023vzw}
Natasha~S. Abrams et~al.
\newblock {Microlensing Discovery and Characterization Efficiency in the Vera C. Rubin Legacy Survey of Space and Time}.
\newblock {\em Astrophys. J. Suppl.}, 276(1):10, 2025.

\bibitem{Perlick:2021aok}
Volker Perlick and Oleg~Yu. Tsupko.
\newblock {Calculating black hole shadows: Review of analytical studies}.
\newblock {\em Phys. Rept.}, 947:1--39, 2022.

\end{thebibliography}

\end{document}